\documentclass{article}
	\usepackage{amsmath,latexsym, graphics,graphicx,showexpl,amsthm,amsfonts,fullpage}
\usepackage[left=1in,top=1in,right=1in,nohead]{geometry}
\usepackage{graphicx}
\usepackage[margin=10pt,font={small},labelfont=bf]{caption}
\usepackage{subcaption}
\usepackage{bbm} 
\usepackage{setspace}
\usepackage{natbib}
\usepackage[titletoc]{appendix}
\usepackage{epstopdf}
\usepackage[pdftex,bookmarks,colorlinks=true,linkcolor=blue]{hyperref}
\usepackage{algpseudocode,algorithm}
\usepackage{authblk}
\usepackage{multicol}
\usepackage{hyperref}
\hypersetup{colorlinks,
	citecolor=blue
}

\evensidemargin=\oddsidemargin
\usepackage{etoolbox}

\newif\ifabbreviation
\pretocmd{\thebibliography}{\abbreviationfalse}{}{}
\AtBeginDocument{\abbreviationtrue}

\begin{document}

\title{Bayesian Design of Experiments \\in the Presence of Nuisance Parameters}
%\title{A sample article title with some additional note\thanksref{t1}}
%\runtitle{Bayesian Design with Nuisance Parameters}
%\thankstext{T1}{A sample additional note to the title.}


	\author{Shirin Golchi$^\dagger$\footnote{Shirin Golchi is the corresponding author and may be contacted at \url{shirin.golchi@mcgill.ca}.} \hspace{35pt} Luke Hagar$^\dagger$ \hspace{25pt} \\ 
$^\dagger$\textit{Biostatistics, McGill University} \\
}

	\date{}

	\maketitle

\begin{abstract}
Design of experiments has traditionally relied on the frequentist hypothesis testing framework where the optimal size of the experiment is specified as the minimum sample size that guarantees a required level of power. Sample size determination may be performed analytically when the test statistic has a known asymptotic sampling distribution and, therefore, the power function is available in analytic form. Bayesian methods have gained popularity in all stages of discovery, namely, design, analysis and decision making. Bayesian decision procedures rely on posterior summaries whose sampling distributions are commonly estimated via Monte Carlo simulations. In the design of scientific studies, the Bayesian approach incorporates uncertainty about the design value(s) instead of conditioning on a single value of the model parameter(s). Accounting for uncertainties in the design value(s) is particularly critical when the model includes nuisance parameters. In this manuscript, we propose methodology that utilizes the large-sample properties of the posterior distribution together with Bayesian additive regression trees (BART) to efficiently obtain the optimal sample size and decision criteria in fixed and adaptive designs. We introduce a fully Bayesian procedure that incorporates the uncertainty associated with the model parameters including the nuisance parameters at the design stage. The proposed approach significantly reduces the computational burden associated with Bayesian design and enables the wide adoption of Bayesian operating characteristics.
\end{abstract}


\section{Introduction}

Sample size determination is an important component of the design of experiments and studies. In classical statistics, sample size estimation follows the frequentist hypothesis testing framework, where a null hypothesis is defined with respect to the value of a quantity of interest. The size of a study is determined as the minimum amount of data required to reject the null hypothesis with a high probability (power) given a hypothesized value under the alternative hypothesis. When the power function of the test is available in analytic form, the sample size may be obtained analytically as is the case for many frequentist test statistics with known asymptotic sampling distributions.

%In addition to the sample size, the power function of a test depends on the ``true'' or hypothesized set of parameters that characterize the data model. While it is often straightforward to hypothesize the value of the parameter of interest, specifying values for the nuisance parameters may appear unsettling in many settings where little information is available about these parameters. Even in presence of information, there is often a significant level of uncertainty which is commonly ignored by treating these parameters as known at the design stage. \cite{}

% \begin{itemize}
% \item  Discuss and cite the non-informative nuisance parameter principle (J.O. Berger, Pericchi and Pererira, Flórez Rivera AF, Esteves LG, Fossaluza V, de Bragança Pereira CA. On the Nuisance Parameter Elimination Principle in Hypothesis Testing. Entropy (Basel). 2024 Jan 29;26(2):117. doi: 10.3390/e26020117. PMID: 38392373; PMCID: PMC10888291.) and its implications for hypothesis testing
% \item (not focused on hypothesis testing but super relevant) Integrated Likelihood Methods for Eliminating Nuisance Parameters,
% James O. Berger, Brunero Liseo and Robert L. Wolpert, Statistical Science 1999 
% \item the mixed test (essentially the Bayesian design)  (Pericchi and Pererira and references therein)
% \item The optimization of weighted sum of error rates (Adaptative significance levels using optimal decision rules: Balancing by weighting the error probabilities, Pericchi and Pererira and references therein)
% \end{itemize}

Bayesian methods have gained popularity in the design and analysis of scientific studies. For example, clinical trials which have traditionally been planned and analyzed using the frequentist framework, have recently adopted Bayesian methods at least partially in various settings \citep{spiegelhalter1994bayesian, spiegelhalter2004bayesian, berry2010bayesian}. For analysis and decision making, the Bayesian approach relies on the posterior distribution and its summaries. A common hybrid Bayesian-frequentist approach is to assess a design that uses Bayesian decision rules with respect to frequentist properties such as frequentist power \cite{berry2010bayesian}. Following \cite{wang2002simulation}, the current practice relies on simulation studies to estimate the sampling distribution of posterior summaries and determine the sample size as well as other design parameters in this hybrid setting.

The term hybrid has also been used to refer to the cases where the Bayesian approach is taken at the design stage while the analysis remains frequentist. Advocated for by many researchers \citep{spiegelhalter_predictive_1986, ohagan_assurance_2005, chuang-stein_sample_2006, gubbiotti2011bayesian, ren_assurance_2014, best_beyond_2024}, this approach incorporates the uncertainty about the parameter of interest via a probability distribution referred to as a design or sampling prior which is distinct from the analysis or fitting prior. The Bayesian approach toward design is particularly advantageous in problems with nuisance parameters as the design prior accommodates the uncertainty associated with these parameters while incorporating a hypothesized set of values for the parameter(s) of interest with their corresponding probabilities. 

The elimination of nuisance parameters in statistical inference has been extensively discussed \cite{KalSpr1970, Basu1977_elimination, BerMor1994_robust, BerLisWol1999_Intlik, RivEstFos2024_nonInformative}. Several proposed approaches rely on integrating the likelihood function \cite{Basu1977_elimination, BerLisWol1999_Intlik}. The Bayesian framework has been noted by several authors as a natural approach to the problem where having assigned priors to all model parameters including the nuisance parameters, inference is made via the marginal posterior distribution of the parameter(s) of interest \cite{Basu1977_elimination, BerMor1994_robust, RivEstFos2024_nonInformative}.

%Of particular relevance is the non-informative nuisance parameter principle \cite{BerWol11988_likprnciple} which, informally, states the irrelevance of the nuisance parameters for inference for the parameter of interest in the case that the likelihood is separable with respect to the two. 

We consider the problem of dealing with nuisance parameters from a design perspective which has received less attention. Recent work of note is that of \citet{RivEstFos2024_nonInformative} who considered the non-informative nuisance parameter principle (NNPP) \cite{BerWol11988_likprnciple} which states the irrelevance of the nuisance parameters for inference about the parameter(s) of interest in the case that the likelihood is separable with respect to the two parameter types. Within this context, they revisited the mixed test approach of \citet{PerPer2016_balancing} which relies on a decision theoretic approach optimizing a weighted sum of errors as a loss function. \citet{RivEstFos2024_nonInformative} showed that the NNPP holds for the mixed test in discrete sample spaces. We are, however, concerned with the general problem of design in presence of (potentially informative) nuisance parameters with a special interest in accommodating Bayesian and decision theoretic approaches similar to that of \citet{PerPer2016_balancing} and its extensions \cite{Calderazzo2020}.

%*** O'Hagan et al. (2005): ``The concept of assurance, although not necessarily this terminology, has been used in work dating back to the 1980s. The idea of computing the Bayesian prior assurance of success in a trial whose data will be analysed in conventional frequentist ways was described by Spiegelhalter and Freedman [6] and is called the ‘hybrid classical–Bayesian’ approach by Spiegelhalter et al. [7, Section 6.5]. It has also been called ‘expected power’ in early work, or ‘average power’ [8]."***

When a design prior is used, the relevant operating characteristics are the integrated power or probability of success, also referred to as assurance or Bayesian assurance. In most realistic models, and especially in the case that a Bayesian analysis is performed, this integration is done numerically via Monte Carlo simulations. Typically, parameter values are generated from the design prior, data are generated from these parameter values, a decision is recorded based on the (frequentist or Bayesian) decision procedure, and this process is repeated to estimate the average power over the design prior and the sampling distribution. A large number of iterations is needed to achieve sufficient precision in the estimates of the operating characteristics.
While there have been previous attempts in reducing the computational overhead required to estimate sampling distributions of posterior summaries \citep{golchi2022estimating, golchi2024estimating, hagar2025scalable, hagar2025design, hagar2025sequential}, efficient Bayesian design in the presence of nuisance parameters remains unaddressed. %Particularly, \citet{golchi2022estimating} proposed a modeling approach relying on Gaussian processes to emulate the sampling distribution across the parameter space. However, this approach relies heavily on a set of Monte Carlo generated instances of the sampling distribution and is highly sensitive to the number of and placement of the parameter values that define the initial simulation scenarios. In addition Gaussian processes do not perform well in cases where the response surface is not sensitive to some of the inputs, i.e., in models with non-informative nuisance parameters.

In this manuscript, we propose using the large-sample properties of the posterior distribution following from the Bernstein-von Mises (BvM) theorem \citep{vaart_asymptotic_1998} together with the asymptotic normality of the maximum likelihood estimator to obtain the sampling distribution of the commonly used posterior summaries in Bayesian testing procedures. This approximate posterior distribution depends on the true values of the nuisance parameters only through a scale parameter which is predicted across the parameter space from a set of initial evaluations generated by simulations and using Bayesian additive regression trees (BART). This approach will result in an analytic power function which can then be efficiently integrated over the design prior.

The fundamental idea for our proposed approach is similar to that of \citet{golchi2022estimating} and \citet{golchi2024estimating}. \citet{golchi2022estimating} emulated the sampling distribution across the parameter space via beta distributions whose parameters are assigned Gaussian process priors. While flexible, this model does not scale to higher dimensional parameter spaces and performs poorly in cases where there is little sensitivity in certain nuisance parameters. \citet{golchi2024estimating} took a more parametric approach that uses the asymptotic properties of the sampling distribution. However, their model does not handle nuisance parameters efficiently.

Other related approaches include \citet{hagar2025scalable} who proposed a method where Bayesian operating characteristics could be assessed across a broad range of sample sizes by estimating the sampling distribution at only two sample sizes. This method was extended by \citet{hagar2025design} to accommodate settings with clustered data and by \citet{hagar2025sequential} to facilitate Bayesian sequential design. Those methods efficiently explore the sample size space for a given design prior; however, they require complete reimplementation for a new design prior. %It would be more efficient to have a method that could use the same simulation results to assess Bayesian operating characteristics with respect to various design priors.

The methods proposed in this manuscript facilitate a decision theoretic approach toward hypothesis testing and design. Particularly, we will showcase the utility of our approach in a decision theoretic setting with a cost function inspired by the mixed test of \citet{Pericchi2016} and its extensions \cite{Calderazzo2020} that aim to unify Bayesian and frequentist approaches to hypothesis testing. 

The remainder of the manuscript is organized as follows. In Section~\ref{Sec:methods}, we formally introduce the problem that we address and present the proposed approach under fixed and group sequential designs. Section~\ref{Sec:example} follows with a stylized example to demonstrate the implementation and numerical performance of the proposed methodology. In Section~\ref{Sec:application}, the proposed methods are applied to a hypothetical Bayesian adaptive design exercise in the context of a real clinical trial, and section~\ref{Sec:discussion} concludes with a discussion.

\section{Methods}
\label{Sec:methods}
\subsection{Background}\label{sec:methods.1}
Consider the design of a study with the objective of testing the following set of one sided hypotheses,
\begin{equation}
\label{eqn:hyp}
	H_0: \psi(\boldsymbol{\theta})\leq \psi_0 \hskip 10pt \text{vs} \hskip 10pt  H_A: \psi(\boldsymbol{\theta})>\psi_0
	\end{equation}
where $\psi(\boldsymbol{\theta})$ is the scalar target of inference given as a function of parameters $\boldsymbol{\theta}$, a $p$-dimensional vector, that define the data model, $\mathbf{Y}\sim f(\mathbf{y}\mid\boldsymbol{\theta})$. We consider the general case where $\boldsymbol{\theta}$ may contain components that do or do not directly contribute to $\psi(\boldsymbol{\theta})$. We refer to the components that do not directly contribute to $\psi(\boldsymbol{\theta})$ as nuisance parameters. However, the nuisance parameters may indirectly inform our knowledge of $\psi(\boldsymbol{\theta})$ as detailed in Section \ref{sec:methods.2}. 

 Considering a parametric Bayesian framework, a prior distribution is assigned to the parameters $\boldsymbol{\theta}$ and inference is made based on the posterior distribution, denoted by $\pi(\boldsymbol{\theta}\mid \mathbf{y}_n)$ from which the posterior distribution for $\psi(\boldsymbol{\theta})$ may be obtained. The notation $\mathbf{y}_n$ represents the observed data from an experiment of size $n$. The decision to reject the null hypothesis is made based on the following summary of the posterior distribution:
\begin{equation}
	\label{eqn:pAlt}
	\tau(\mathbf{y}_n) = P\left(\psi(\boldsymbol{\theta})>\psi_0\mid \mathbf{y}_n\right). 
\end{equation}
For example, a decision of ``success'' is made if $\tau(\mathbf{y}_n)>u$ for some prespecified probability threshold $u$. Although $\tau$ is obtained by conditioning on one set of observed data  at the analysis/decision stage, at the design stage we are interested in the error rates associated with decision processes defined based on $\tau$ over the distribution of the data under the assumed data model, i.e., the sampling distribution. This is the reason for denoting the decision statistic explicitly as a function of $\mathbf{Y}_n$, i.e., the random data arising from an experiment of size $n$. The sampling properties of the decision procedure depend on the sampling distribution of $\tau(\mathbf{Y}_n)$. 

Under the frequentist design (hybrid approach) where a set of fixed parameter values is hypothesized, we need to obtain the sampling distribution of $\tau(\mathbf{Y}_n)$ given $\boldsymbol{\theta}=\boldsymbol{\theta}^*$, i.e., $f(\tau(\mathbf{Y}_n)\mid \boldsymbol{\theta}=\boldsymbol{\theta}^*)$. From this sampling distribution, the power function and the frequentist operating characteristics can be computed:
\begin{align}
\label{eq:power}
    \Pi(\boldsymbol{\theta}) &= P(\tau(\mathbf{Y}_n)>u | \boldsymbol{\theta})\\ 
    &=\int_u^1 f(\tau(\mathbf{y}_n)\mid \boldsymbol{\theta})d\tau(\mathbf{y}_n). \nonumber
\end{align}
In the Bayesian design framework, the uncertainty about the hypothesized parameter values is expressed via a design prior distribution $\pi_d(\boldsymbol{\theta})$. Bayesian operating characteristics are then obtained by integrating the power function over this design prior,
\begin{align}
    \Pi_B &= \int \Pi(\boldsymbol{\theta})\pi_d(\boldsymbol{\theta})d\boldsymbol{\theta}\\ \nonumber
    &=\int\left\{\int_u^1 f(\tau(\mathbf{y}_n)\mid \boldsymbol{\theta})d\tau(\mathbf{y}_n)\right\}\pi_d(\boldsymbol{\theta})d\boldsymbol{\theta}\\ \nonumber
    &=\int\int_u^1 f(\tau(\mathbf{y}_n)\mid \boldsymbol{\theta})\pi_d(\boldsymbol{\theta})d\tau(\mathbf{y}_n)d\boldsymbol{\theta}.\nonumber
\end{align}
In practice, the above integrals are approximated using Monte Carlo simulations. Considering that various design configurations need to be explored, these simulations can be computationally intensive, especially when the evaluation of $\tau$ requires additional computations such as Markov chain Monte Carlo (MCMC) sampling or other posterior approximation approaches. Estimating the Bayesian operating characteristics of the form $\Pi_B$ also significantly adds to the computational burden as they require numerically evaluating a double integral in the simplest case where $\boldsymbol{\theta}$ is a scalar, but a multiple integral for higher dimensional models. The latter is the focus of this manuscript. We will introduce methodology for approximating $f(\tau(\mathbf{y}_n)\mid \boldsymbol{\theta})$ across the parameter space $\Theta$ which allows for evaluation of $\Pi_B$ at a significantly lower computational cost than any existing method.

\subsection{Approximating the sampling distribution using BART}\label{sec:methods.2}

As introduced in Section \ref{sec:methods.1}, $\psi = \psi(\boldsymbol{\theta})$ is the quantity of interest with respect to which the hypotheses are formulated as in (\ref{eqn:hyp}). Consider $\boldsymbol{\eta}$ as the subset of parameters in $\boldsymbol{\theta}$ that do not directly contribute to $ \psi(\boldsymbol{\theta})$. We refer to  $\boldsymbol{\eta}$ as nuisance parameters that are needed for fully specifying the data model. 

Suppose that a unique maximum likelihood estimator (MLE) obtained from an experiment of size $n$ exists for $\boldsymbol{\theta}$, denoted by $\hat{\boldsymbol{\theta}}_n$. The corresponding MLEs $\hat{\psi}_n$ and $\hat{\boldsymbol{\eta}}_n$ can be defined as functions of $\hat{\boldsymbol{\theta}}_n$.  Under weak regularity conditions (compact parameter space, continuity and identifiablity) the symmetry below follows from the Bernstein-von Mises theorem \citep{vaart_asymptotic_1998} as well as the asymptotic normality and consistency of the MLE \citep{lehmann1998theory}:
\begin{equation}
	\label{eqn:von-Mises}
\psi\mid \hat{\psi}_n \overset{.}{\sim} \mathcal{N}\left(\hat{\psi}_n, \frac{1}{n}\frac{\partial \psi}{\partial\boldsymbol{\theta}}\bigg|_{\boldsymbol{\theta} = \hat{\boldsymbol{\theta}}_n}^{\text{T}}I^{-1}_{\hat{\boldsymbol{\theta}}_n}\frac{\partial \psi}{\partial\boldsymbol{\theta}}\bigg|_{\boldsymbol{\theta} = \hat{\boldsymbol{\theta}}_n}\right), \end{equation}

\begin{equation}
	\label{eqn:CLT}
\hat{\psi}_n\mid \psi = \psi^* \overset{.}{\sim} \mathcal{N}\left( \psi^*, \frac{1}{n}\frac{\partial \psi}{\partial\boldsymbol{\theta}}\bigg|_{\boldsymbol{\theta} = \boldsymbol{\theta}^*}^{\text{T}}I^{-1}_{\boldsymbol{\theta}^*}\frac{\partial \psi}{\partial\boldsymbol{\theta}}\bigg|_{\boldsymbol{\theta} = \boldsymbol{\theta}^*}\right).
\end{equation}
where $\psi^* = \psi(\theta^*)$, and $I^{-1}_{\hat{\boldsymbol{\theta}}_n}$ and $I^{-1}_{\boldsymbol{\theta}^*}$ denote the inverse Fisher information evaluated at $\hat{\boldsymbol{\theta}}_n$ and $\boldsymbol{\theta}^*$, respectively. While $\boldsymbol{\eta}$ does not directly impact the value of $\psi$, the nuisance parameters may impact (\ref{eqn:von-Mises}) and (\ref{eqn:CLT}) through the Fisher information. 

To simplify the notation, let $\hat{\lambda}/n$ be the variance in (\ref{eqn:von-Mises}) and $\lambda^*/n$ be the variance in (\ref{eqn:CLT}). 
Then (\ref{eqn:von-Mises}) and (\ref{eqn:CLT}) result in the following approximate model for the decision statistic $\tau$, 
\begin{align*}
	\tau_n &\approx \Phi\left(\frac{\sqrt{n}}{\hat{\lambda}}( \hat{\psi}_n - \psi^*) + \frac{\sqrt{n}}{\hat{\lambda}}(\psi^*-\psi_0)\right) \\
    &=\Phi\left(\frac{\lambda^*}{\hat{\lambda}}\frac{\sqrt{n}}{\lambda^*}( \hat{\psi}_n - \psi^*) + \frac{\sqrt{n}}{\hat{\lambda}}(\psi^*-\psi_0)\right) \\
    &=\Phi\left(\frac{\lambda^*}{\hat{\lambda}}\epsilon + \frac{\sqrt{n}}{\hat{\lambda}}(\psi^*-\psi_0)\right)
	\end{align*}
where $\epsilon \sim N(0,1)$.

Therefore, a probit transformation of the decision statistic is approximately linear in $\sqrt{n}( \psi^*- \psi_0)$ which does not depend on the nuisance parameters: 
 \begin{align*}
	\gamma_n = \Phi^{-1}(\tau_n) &\approx  \frac{\sqrt{n}}{\hat{\lambda}}( \psi^*- \psi_0) + \frac{\lambda^*}{\hat{\lambda}}\epsilon.
\end{align*} 
For large enough sample sizes, $\hat{\lambda} \approx \lambda^* = \lambda$, simplifying the above expression further to 
 \begin{align*}
	\gamma_n = \Phi^{-1}(\tau) &\approx  \frac{\sqrt{n}}{\lambda}( \psi^*- \psi_0) + \epsilon.
\end{align*} 
As $\lambda$ is the only term that depends on the (unspecified) model parameters, to learn the approximate sampling distribution of $\tau_n$, we need to the learn $\lambda$ across the parameter space. 

In simpler models where the Fisher information can be obtained, the dependence of $\lambda$ on the nuisance parameters is exactly known. Particularly, in cases where the likelihood is tractable we can identify the informative nuisance parameters as those that contribute to the Fisher information. %Specifically, in cases where the likelihood can be factored in terms of $\psi$ and nuisance parameters, the sampling distribution would not be sensitive to the nuisance parameters. %is this clear enough?

However, our goal is to propose an automated process that applies to any analysis model regardless of the complexity as long as the regularity conditions are satisfied. Therefore, instead of relying on obtaining the Fisher information, we propose generating a set of estimates for $\lambda$, obtained from the posterior variance at an initial set of values $\boldsymbol{\theta}\in \boldsymbol{\Theta}$, to be used as ``training data'' for a prediction model that can be used to generate Fisher information estimates for any parameter values across the parameter space. 

This approach is fundamentally similar to that of \cite{golchi2022estimating} who used a Gaussian process to learn certain parameters of an approximate sampling distribution. However, Gaussian processes do not perform well in cases where the sampling distribution is not sensitive to a subset of inputs. Therefore, we propose using Bayesian additive regression trees, which were shown to have high predictive performance with a relatively large number of predictors when only a few may in fact be important \citep{BART_Chipman_2010}. The model for $\lambda$ is presented as a sum of $m$ trees,
\begin{equation}
\label{eqn:bart}
\lambda = \sum_{j=1}^m g(\boldsymbol{\theta}; T_j, M_j) + \epsilon_{\lambda}, \hskip 20pt \epsilon_{\lambda}\sim N(0,\sigma^2)
\end{equation}
where each $T_j$ is a binary regression tree whose terminal node parameters are denoted by $M_j$. Prior distributions are then specified on $(T_j, M_j)$ and $\sigma$. Particularly we adopt the Soft BART approach of \citet{SoftBART2018} that can adapt to unknown levels of smoothness and sparsity via probabilistic branching implemented through the \texttt{SoftBART} R package.

This model is then trained over a set of Monte Carlo generated values of $\lambda$ at selected points throughout a subregion of the parameter space, $\boldsymbol{\Theta}_d \in \boldsymbol{\Theta}$, where the design prior, $\pi_d(\boldsymbol{\theta})$, has non-negligible density. Predictions of $\lambda$ are then obtained for any $\boldsymbol{\theta} \in \boldsymbol{\Theta}$ which will fully specify the distribution of $\tau$. Frequentist operating characteristics are then immediately obtained as the summaries of this distribution for any given $\boldsymbol{\theta}$ and Bayesian operating characteristics are computed by numerically integrating these over the design prior. Particularly, the power function defined in (\ref{eq:power}) is estimated by
\begin{align}
\label{eqn:power}
\Pi(\boldsymbol{\theta}) \approx 1 - \Phi\left(\Phi^{-1}(u)- \frac{\sqrt{n}}{\hat{\lambda}(\boldsymbol{\theta})}\left(\psi(\boldsymbol{\theta})-\psi(\boldsymbol{\theta}_0)\right)\right),
\end{align}
and the Bayesian integrated power by
\begin{align}
\label{eqn:assurance}
\Pi_B \approx \int\left\{1 - \Phi\left(\Phi^{-1}(u)- \frac{\sqrt{n}}{\hat{\lambda}(\boldsymbol{\theta})}\left(\psi(\boldsymbol{\theta})-\psi(\boldsymbol{\theta}_0)\right)\right)\right\}\pi_d(\boldsymbol{\theta})d\boldsymbol{\theta}.
\end{align}
Note that since BART provides samples from the posterior distribution of $\lambda(\boldsymbol{\theta})$, the uncertainty associated with the estimation of $\lambda(\boldsymbol{\theta})$ may be incorporated into the estimates of the operating characteristics and represented by credible intervals. Since the proposed approach relies on BART predictions of the Fisher information and the normal approximation given by the Bernstein-von Mises theorem, we will refer to it as the BART-BvM approach. 

\subsection{Bayesian group sequential designs}
 In group sequential designs (GSDs), the experiment is divided into stages with an analysis and decision at the end of each stage. Various stopping decisions may be considered in group sequential designs. The decision boundaries may vary across the stages, which allows for tuning the design to attain desired error probabilities associated with each decision point and the overall decision procedure. This practice allows for flexibility, increased efficiency, and potential sample size savings at the cost of complicating the design. In classical GSDs, decision boundaries and other design parameters such as the number and spacing of interim analyses are specified using group sequential theory that relies on the properties of frequentist tests.

 Bayesian GSDs (also referred to as Bayesian adaptive designs) rely on decision summaries of the form defined in (\ref{eqn:pAlt}). The design parameters in Bayesian GSDs are mainly selected through Monte Carlo simulations. Given the flexibility of these designs, the number of possible scenarios to be explored in these simulation studies is often large.
 
The method described in Section \ref{sec:methods.1} can be extended to Bayesian GSDs.  We consider GSDs with $T$ planned analyses indexed by $t \in \{1, \dots, T\}$. At analysis $t$, the $n_t$ cumulative observations comprise the data, $\mathbf{y}_{n_t}$. All observations from previous stages are retained in $\mathbf{y}_{n_t}$ for subsequent analyses. The posterior probability in (\ref{eqn:pAlt}) is considered at the end of each stage, prompting a vector of $T$ posterior summaries:
  \begin{equation}\label{eq:pp}
           \boldsymbol{\tau}  = 
           \begin{bmatrix}
           \tau(\mathbf{y}_{n_1}) \\
           \vdots \\
           \tau(\mathbf{y}_{n_T}) 
         \end{bmatrix} = \begin{bmatrix}
           P\left(\psi(\boldsymbol{\theta})>\psi_0\mid \mathbf{y}_{n_1}\right) \\
           \vdots \\
           P\left(\psi(\boldsymbol{\theta})>\psi_0\mid \mathbf{y}_{n_T}\right) 
         \end{bmatrix}.
\end{equation} 
For instance, the experiment may stop for ``success'' after stage $t$ if $\tau(\mathbf{y}_{n_t}) > u_t$, a stage-specific probability threshold.

We emphasize that data from all $T$ stages of a GSD -- and the corresponding posterior summaries in (\ref{eq:pp}) -- are not fully observed in a given experiment if it stops early. As such, we describe how to estimate stopping probabilities at each stage of the experiment by considering the large-sample theoretical results about the sampling distribution of $\boldsymbol{\tau}$ with respect to the \emph{complete observable} data $\{\mathbf{Y}_{n_t}\}_{t=1}^T$ across all $T$ analyses. 

We let $\hat{\boldsymbol{\psi}} = (\hat{\psi}_{n_1}, \dots, \hat{\psi}_{n_T})$ be the joint MLE for $\psi$ across all analyses. The results below follow from the BvM theorem and the joint canonical distribution \citep{jennison1999group} inherent to group sequential theory:
\begin{equation}
	\label{eqn:von-Mises.T}
\psi\mid \hat{\psi}_{n_t} \overset{.}{\sim} \mathcal{N}\left(\hat{\psi}_{n_t}, \frac{1}{n_t}\frac{\partial \psi}{\partial\boldsymbol{\theta}}\bigg|_{\boldsymbol{\theta} = \hat{\boldsymbol{\theta}}_{n_t}}^{\text{T}}I^{-1}_{\hat{\boldsymbol{\theta}}_{n_t}}\frac{\partial \psi}{\partial\boldsymbol{\theta}}\bigg|_{\boldsymbol{\theta} = \hat{\boldsymbol{\theta}}_{n_t}}\right), \end{equation}

\begin{equation}
	\label{eqn:CLT.T}
\hat{\boldsymbol{\psi}}\mid \psi = \psi^* \overset{.}{\sim} \mathcal{N}\left( \psi^* \times \mathbf{1}_T, \lambda^* \times {\mathbf{C}}\right),
\end{equation}
where the $(j,k)$-entry of the matrix ${\mathbf{C}}$ is $\min\{n_j^{-1}, n_k^{-1}\}$ for $j, k \in \{1, \dots, T\}$. We let $\hat{\lambda}_t/n_t$ be the variance of the distribution in (\ref{eqn:von-Mises.T}). It is reasonable to assume that $\lambda$ remains constant through the stages of the experiment under standard regularity conditions. For large enough sample sizes, we have that $\hat{\lambda}_1 \approx \dots \approx \hat{\lambda}_T \approx \lambda^* = \lambda$.

Following the logic from Section \ref{sec:methods.2}, the probit transformation of the decision statistic at analysis $t$ is such that
 \begin{align*}
	\gamma_{n_t} = \Phi^{-1}(\tau_{n_t}) &\approx  \frac{\sqrt{n_t}}{\lambda}( \psi^*- \psi_0) + \epsilon_t,
\end{align*} 
 where $\epsilon_t \sim \mathcal{N}(0,1)$. As for the fixed design case, the expression above only depends on the nuisance parameters through $\lambda$. We obtain the approximate joint sampling distribution of $\boldsymbol{\gamma} = (\gamma_{n_1}, \dots, \gamma_{n_T})$ by considering the covariance between $\epsilon_j$ and $\epsilon_k$ for $1 \le j \le k \le T$. By the joint canonical distribution \citep{jennison1999group}, we have that
  \begin{align*}
	\text{Cov}(\epsilon_j, \epsilon_k) &= \text{Cov}\left(\dfrac{\sqrt{n_j}(\hat{\psi}_{n_j} - \psi^*)}{\lambda}, \dfrac{\sqrt{n_k}(\hat{\psi}_{n_k} - \psi^*)}{\lambda}\right) \\
    &= \dfrac{\sqrt{n_jn_k}}{\lambda^2}\text{Cov}(\hat{\psi}_{n_j}, \hat{\psi}_{n_k}) \\ &= \dfrac{\sqrt{n_jn_k}}{\lambda^2}\text{Var}(\hat{\psi}_{n_k}) \\ &= \dfrac{\sqrt{n_jn_k}}{\lambda^2}\dfrac{\lambda^2}{n_k} = \sqrt{n_j/n_k}
\end{align*} 
 
% for analysis $t$, where $\epsilon_t \sim N(0,1)$. Moreover,
%  \begin{align*}
% 	\gamma_2 = \Phi^{-1}(\tau_2) &\approx  \frac{\sqrt{n_2}}{\lambda}( \psi^*- \psi_0) + \epsilon_2
% \end{align*} 
% for the second analysis, where $\epsilon_2 \sim N(0,1)$. Once $\lambda$ is estimated for a particular design scenario, we can get, for instance, power for a particular sample size by integrating with respect to a multivariate normal distribution. To calculate power, we can work with $\gamma_1, \gamma_2, \dots$ if we consider the probits of the decision thresholds. 

% By the canonical form of joint sampling distributions, we can determine the level of correlation between $\epsilon_1$ and $\epsilon_2$. Since $\epsilon_1$ and $\epsilon_2$ have unit variance, their correlation is equal to the covariance. This covariance is

% The equivalence of the second and third line follows from standard results about the joint canonical distribution. In general, the correlation between the error terms should be $\sqrt{n_{k_1}/n_{k_2}}$, where $k_1 < k_2$ are the indices for the two relevant analyses. So this result can be extended to multiple interim analyses.

We therefore define the following multivariate normal random variable that represents the joint decision statistic  for an experiment with $T$ analyses:  
\begin{equation}\label{eq:joint.gamma}
\boldsymbol{\gamma} \overset{.}{\sim} \mathcal{N}\left(\dfrac{(\psi^* - \psi_0)}{\hat{\lambda}}\begin{pmatrix}
\sqrt{n_1}\\
\sqrt{n_2}\\
\vdots \\
\sqrt{n_T}
\end{pmatrix}, \begin{bmatrix}
1 & \sqrt{n_1/n_2} & \dots & \sqrt{n_1/n_T}\\
\sqrt{n_1/n_2} & 1 & \dots & \sqrt{n_2/n_T}\\
\vdots & \vdots & \ddots & \vdots \\
\sqrt{n_1/n_T} & \sqrt{n_2/n_T} & \dots & 1
\end{bmatrix}\right).
\end{equation}
This joint distribution of the decision statistics at all $T$ stages enables the computation of cumulative probabilities of stopping up to each analysis as well as power. For example, 
the probability that the experiment stops for success at analysis $t \le T$ is 
$$P(\gamma_1 < \Phi^{-1}(u_1), \gamma_2 < \Phi^{-1}(u_2), \dots, \gamma_t \ge \Phi^{-1}(u_t)),$$
which can be derived from the above multivariate normal sampling distribution by appropriate conditioning.

The process to estimate $\lambda(\boldsymbol{\theta})$ for GSDs using BART is the same as in Section \ref{sec:methods.2}. We need only estimate $\lambda$ using Monte Carlo simulations conducted for an initial set of $\boldsymbol{\theta} \in \boldsymbol{\Theta}$ values at a single sample size $n$. The differing cumulative sample sizes over all stages of the experiment can be explicitly accounted for via (\ref{eq:joint.gamma}).

% A process to calculate Bayesian assurance for a sequential design would involve (i) estimating $\lambda$ for many different $\boldsymbol{\theta}$ values via BART, (ii) integrating with respect to a multivariate normal distribution for $\epsilon_1, \dots, \epsilon_T$ to get conditional power for each $\boldsymbol{\theta}$ value, and (iii) averaging over these conditional powers to get Bayesian assurance. Does this make sense?

For a fixed value of the parameters $\boldsymbol{\theta}$, the probability of stopping for efficacy at analysis $t$ can be estimated using BART-BvM as
\begin{align}
\label{eqn:power.gsd}
\Pi^t(\boldsymbol{\theta}) \approx \int_{-\infty}^{\Phi^{-1}(u_1)}\int_{-\infty}^{\Phi^{-1}(u_2)}\dots \int_{\Phi^{-1}(u_t)}^{\infty} \phi_t(\gamma_1, \dots, \gamma_t; \psi^*, \psi_0, \{n_k\}_{k=1}^t, \hat{\lambda}(\boldsymbol{\theta}))d\gamma_t \dots d\gamma_1,
\end{align}
where $\phi_t(~\cdot~; \psi^*, \psi_0, \{n_k\}_{k=1}^t, \hat{\lambda}(\boldsymbol{\theta}))$ is the probability density function of the multivariate normal distribution in (\ref{eq:joint.gamma}) parameterized by $\psi^*, \psi_0, \{n_k\}_{k=1}^t$, and $\hat{\lambda}(\boldsymbol{\theta})$. 
Given a design prior $\pi_d(\boldsymbol{\theta})$, we can numerically estimate the corresponding integrated probability of stopping as
\begin{align}
\label{eqn:assurance.gsd}
\Pi^t_B \approx \int\Pi^t(\boldsymbol{\theta})\pi_d(\boldsymbol{\theta})d\boldsymbol{\theta}.
\end{align}

\section{Implementation and numerical assessment}
\label{Sec:example}
In this section, we demonstrate the implementation and performance of the BART-BvM approach in the context of a stylized example with a simple multi-parameter analysis model and a fixed design.  Specifically, we assume the following logistic regression model for the binary outcome $Y$ in terms of a binary treatment assignment $A$ (e.g., placebo vs treatment) and a binary covariate $x$, which defines two subgroups with potentially heterogeneous responses to the treatment:
\begin{align*}
&Y\sim \text{Binom}(1, p)\\
&\text{logit}(p) = \beta_0 + \beta_1 x + (\psi_0 + \psi_1x)A.
\end{align*}
As an analysis prior, we independently assign weakly informative $\mathcal{N}(0, 2.5^2)$ distributions to all the parameters. The parameters of interest are $\psi_0$ and $\psi_1$ that represent the (conditional) treatment effect in each of the subgroups. The parameters $\beta_0$ and $\beta_1$ determine the probabilities of response in the two subgroups in absence of the treatment. We will focus on $\psi_0$ and for demonstrative purposes treat all the other parameters as nuisance parameters that can potentially contribute to the Fisher information for $\psi_0$.  Although,  it is straightforward to see in this simple model that the parameter that contributes most to the Fisher information and therefore the sampling distribution of the test concerning $\psi_0$ is $\beta_0$. 

We consider the design prior as the following independent normal distributions over the model parameters: 
\begin{align*}
\beta_0 \sim \mathcal{N}(0, 0.6^2), \hskip 5pt
\beta_1 \sim \mathcal{N}(0, 0.075^2),\\
\psi_0 \sim \mathcal{N}(0.3, 0.15^2), \hskip 5pt 
\psi_1 \sim \mathcal{N}(0, 0.05^2).
\end{align*}
We specify the training set as a Latin hypercube design \cite{LHD1979} of size 40 over the 4-dimensional hypercube defined by the range for each parameter as two standard deviations away from the mean of the design prior,
\begin{equation*}
\boldsymbol{\Theta}_d = [-1.2,1.2]\times [-0.15,0.15]\times [0,0.6]\times [-0.1,0.1].
\end{equation*}
Estimates of $\lambda$ are then obtained over the training set by simulating the sampling distribution of the posterior mean for a given sample size (here $n = 500$), and calculating $\sqrt{n}$ times the standard deviation of this sampling distribution.

\begin{figure}[!tb] \centering 
	\begin{subfigure}[b]{0.47\textwidth}
		\centering
		\includegraphics[width=\textwidth]{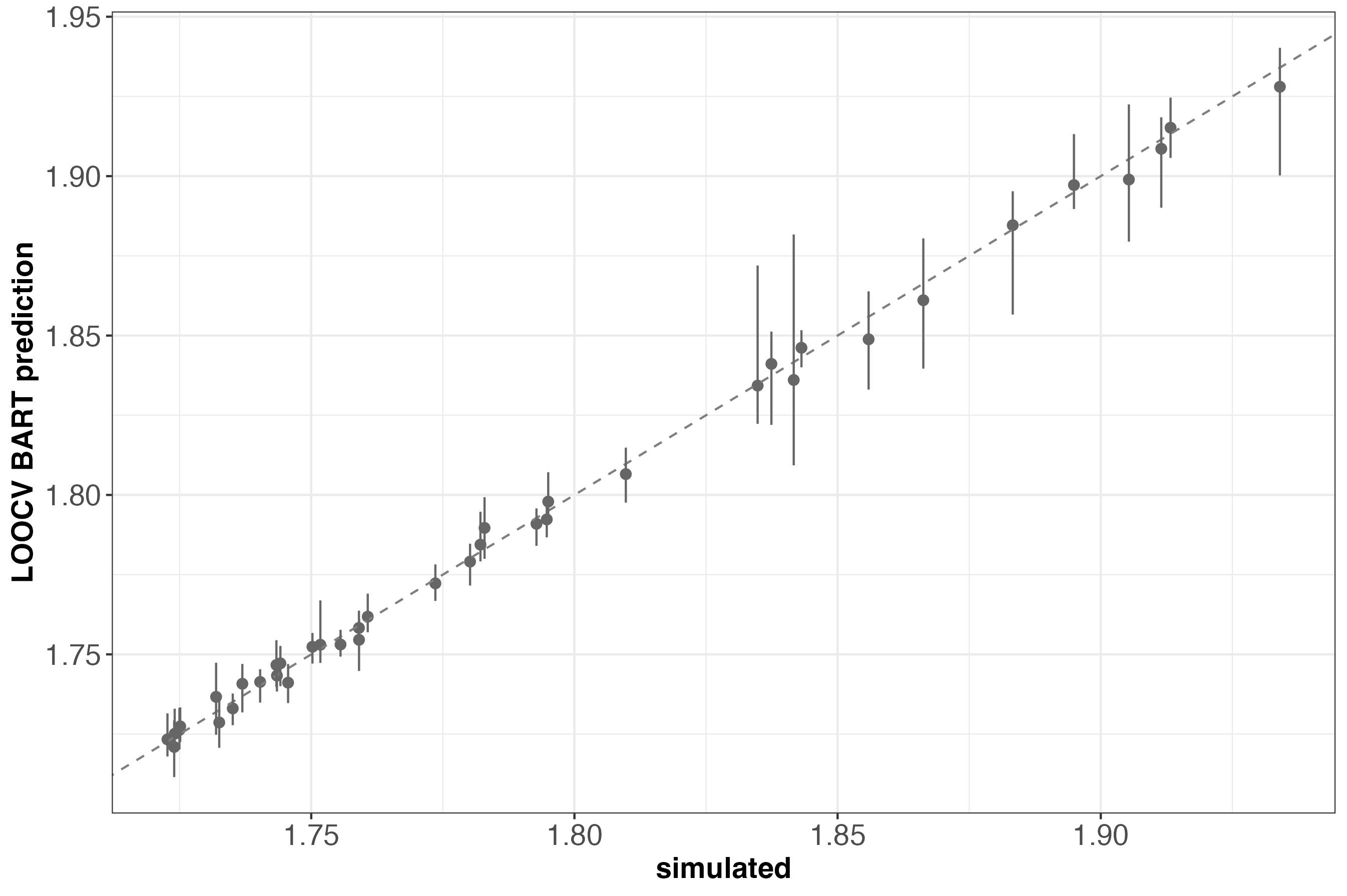}
		\caption{}
		\label{fig:loocvlambda}
	\end{subfigure}
    	\begin{subfigure}[b]{0.47\textwidth}
		\centering
		\includegraphics[width=\textwidth]{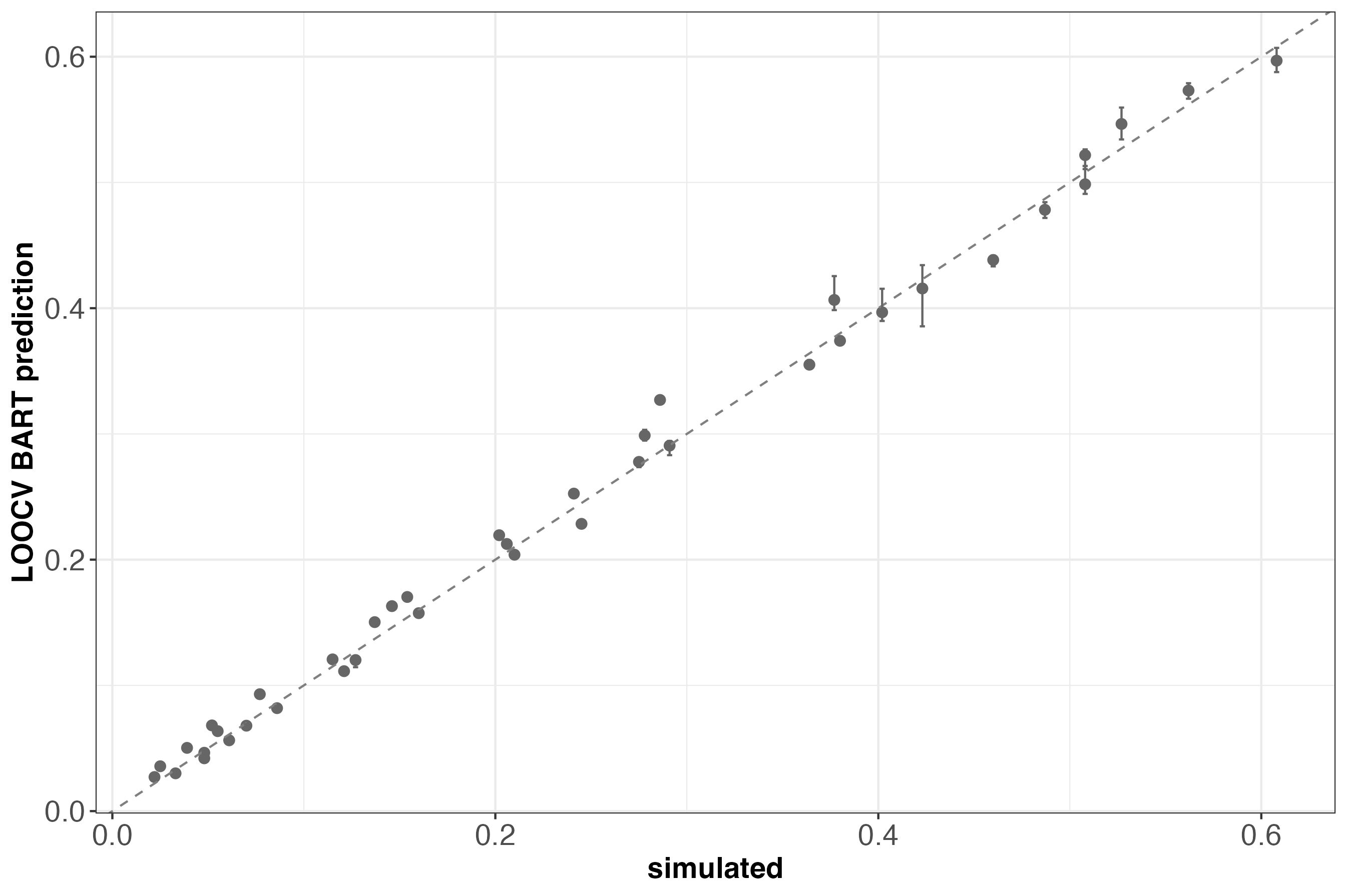}
		\caption{}
		\label{fig:loocvpower}
	\end{subfigure}
\caption{(a) LOOCV BART predictions of $\lambda$ versus simulated values on the log scale and (b) LOOCV BART-BvM estimates of power versus Monte Carlo estimates of power for the 40 samples across $\Theta_d$ together with 95\% credible intervals.} 
\end{figure}

The probabilistic BART model is trained over the set of simulated values for $\hat{\lambda}$ on the log scale to provide predictions of $\lambda$ across $\Theta_d$. Figure~\ref{fig:loocvlambda} shows the leave-one-out cross-validated (LOOCV) predictions generated from the model against those generated by Monte Carlo simulations. Figure~\ref{fig:loocvpower} shows the estimates of power obtained from the predicted $\lambda$ values and the BvM approximation. These figures verify the acceptable performance of the BART prediction of the Fisher information and the satisfactory quality of the approximation to power in (\ref{eqn:power}) compared to corresponding estimates obtained via Monte Carlo. 

The trained BART model can then be used to generate estimates of $\lambda$ for any point across $\Theta_d$ to compute the power integrated over the design distribution of the nuisance parameters. These estimates are obtained by plugging $\hat{\lambda}$ into (\ref{eqn:power}) for a Monte Carlo sample drawn from the design prior and computing the average over the sampled values. Note that integrating these estimates further over the design prior of $\psi_0$ will result in estimates of Bayesian assurance. Our approach enables the instant estimation of integrated power curves for any sample size and any set of parameters; although, due to the asymptotic nature of the approximation in (\ref{eqn:power}), it is expected to be less satisfactory for smaller sample sizes. 

\begin{figure}[!tb] \centering 	
		\centering
		\includegraphics[width=0.75\textwidth]{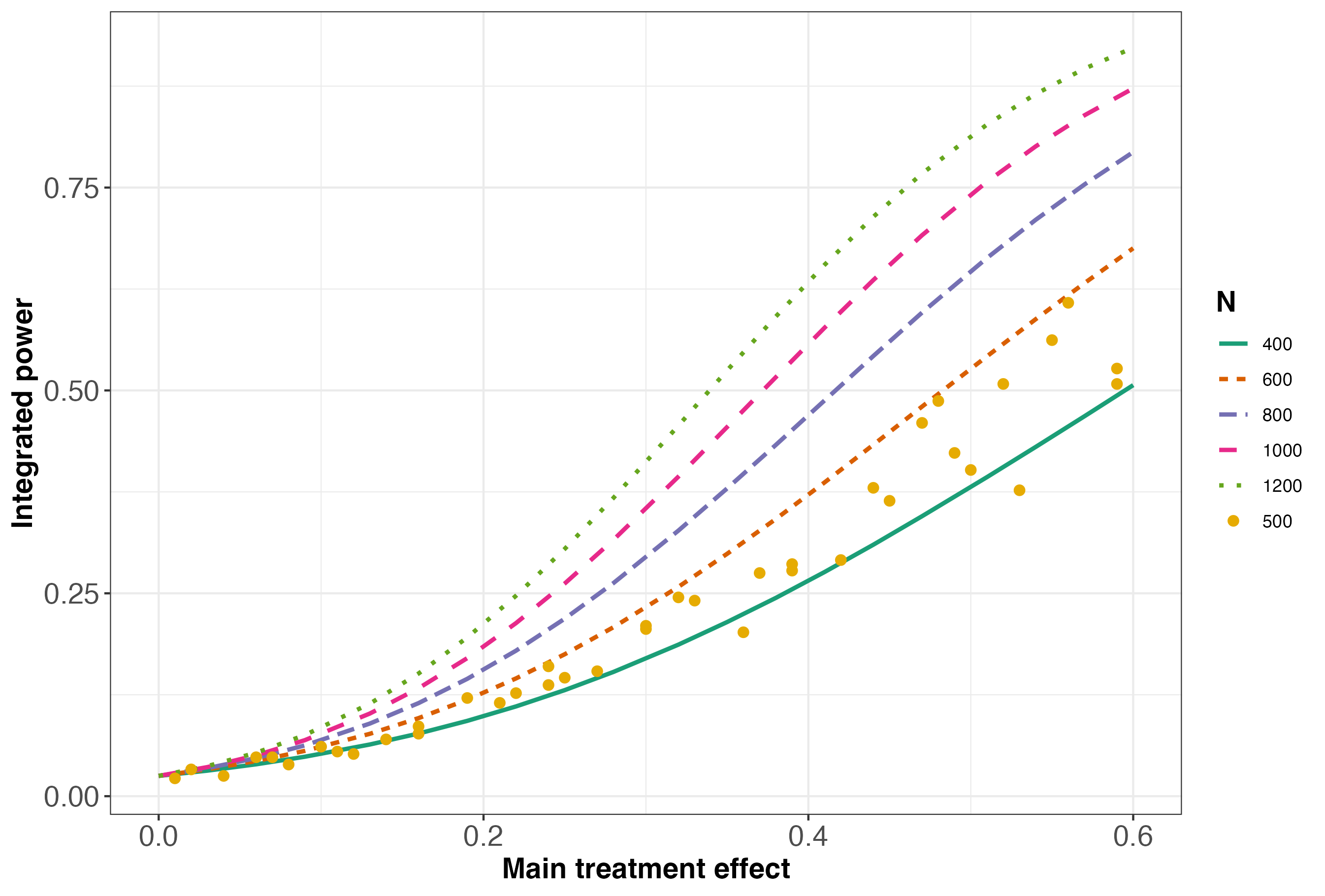}
\caption{\label{fig:IntPowerVsN}The curves are BART-BvM estimates of power integrated over the nuisance parameters versus the conditional main effect of treatment for a range of sample sizes. The dots are the Monte Carlo estimates with $N=500$.} 
\end{figure}

To illustrate the utility of the proposed approach in efficient design and sample size assessment while accommodating uncertainty with respect to the nuisance parameters, we obtain BART-BvM estimates of power over a grid of size $20^4$ throughout $\Theta_d$. Then we integrate over the design prior for the parameters $\beta_0$, $\beta_1$ and $\psi_1$ to obtain integrated power curves against values of $\psi_0$ for a range of increasing sample sizes. Figure~\ref{fig:IntPowerVsN} shows the BART-BvM integrated power curves as well as the Monte Carlo estimates obtained at the 40 points in the training set (yellow dots) for a sample size of 500.

\section{Application to Bayesian group sequential designs}
\label{Sec:application}
In this section, we demonstrate the implementation and applicability of the proposed approach in the context of Bayesian group sequential designs. As a substantive motivation, we use the Canadian placebo-controlled randomized trial of tecovirimat in non-hospitalized patients with Mpox (PLATINUM-CAN) \citep{klien2024tecovirimat}. This trial employs a fixed randomized design with frequentist analysis. The main goal of the trial is to establish that the antiviral drug tecovirimat reduces the duration of illness associated with Mpox infection. The trial's primary outcome is the time to lesion resolution, defined as the first day after randomization on which all Mpox lesions are completely resolved. The impact of tecovirimat on the time to lesion resolution will be evaluated in comparison to a placebo, with balanced randomization to the treatment and control arms. While we use the substantive research question of this clinical trial as a motivating example, we propose a Bayesian analysis model and a Bayesian group sequential design framework to showcase the methods of this article.

{\bf Bayesian analysis model} We define the baseline ``hazard'' of experiencing lesion resolution as a piecewise constant function and let $s$ be the time since randomization in days. We allow the baseline hazard $h_0(s)$ to change with the week of observation such that $h_0(s)$ is $h_1$ if $0 \le s \le 7$, $h_2$ if $7 < s \le 14$, $h_3$ if $14 < s \le 21$, and $h_4$ if $s > 21$. We define the censoring process such that 
for the clinical visit on day $7j$ with $j = 1, \dots, 4$, there is a probability of $\kappa$ that each patient who has not already experienced lesion resolution drops out of the study. The lesion resolution times for those patients are right censored at the time of their last-attended visit (i.e., day $7(j-1)$). The resolution times of all patients who have not experienced lesion resolution by the end of the 28-day follow-up period are right censored at 28 days.

The hazard function for patient $i$ is
    \begin{equation*}\label{eq:hazard2}
    h_i(s~|~A_{i}; \boldsymbol{\theta}) = h_0(s)\exp(\beta A_{i}),
    \end{equation*}
    where $A=1$ indicates being randomized into the tecovirimat arm. We introduce a binary variable $\delta$ such that $\delta_i = 1$ if patient $i$'s recorded time is a lesion resolution time and not a censoring time.
    The likelihood function for the $n$ patients' data is
    $$\prod_{i=1}^n h_i(s_i~|~A_{i}; \boldsymbol{\theta})^{\delta_i}\exp\left(-\int_{0}^{s_i}h_i(s_i~|~A_{i}; \boldsymbol{\theta})ds\right).$$
    The model for Bayesian analysis is completed by independently assigning the following diffuse analysis priors:
    \begin{align*}
&\beta \sim \mathcal{N}(0,10^2), \hskip 10pt
\log(h_j) \sim \mathcal{N}(0,10^2) ~\text{for} ~ j = 1, 2, 3, 4.
\end{align*}

    The target of inference $\psi(\boldsymbol{\theta})$ for this trial is the population-level rate ratio of experiencing lesion resolution: $h(s~|~A = 1; \boldsymbol{\theta})/h(s~|~A = 0; \boldsymbol{\theta}) = \exp(\beta)$. %That is, the marginal and conditional estimands coincide in this case.
    Therefore, the model parameters include the baseline ``hazards'' of experiencing lesion resolution $\boldsymbol{h} = (h_1, h_2, h_3, h_4) \in \mathbb{R}^4_+$, the regression coefficient $\beta \in \mathbb{R}$, and the censoring parameter $\kappa \in [0,1]$ resulting in a six-dimensional parameter space with five nuisance parameters. 

   % These are the parameter values that are closest to what we used in the sequential paper with the more complex example: $\boldsymbol{h} = (0.055, 0.095, 0.04, 0.0200)$, $\zeta = 1/2$, and $\kappa = 0.0250$. For the probability model $\Psi_0$ used to consider the type I error rate, the regression coefficient $\beta = 0$ was used. Our main $\Psi_1$ model used to consider power was defined with $\beta = \log(1.3)$. Coefficients of $\beta = \log(1.4)$ and $\beta = \log(1.5)$ were also considered for sensitivity analysis.
    
{\bf Design prior} The following independent design prior represents the hypothetical prior data and assumptions regarding the model parameters at the design stage,
\begin{align*}
&\log(h_1) \sim \mathcal{N}(\log(0.055), 0.15^2), \hskip 10pt
\log(h_2) \sim \mathcal{N}(\log(0.095), 0.15^2),\\
&\log(h_3) \sim \mathcal{N}(\log(0.040), 0.15^2), \hskip 10pt 
\log(h_4) \sim \mathcal{N}(\log(0.020), 0.15^2),
\\
&\beta \sim \mathcal{N}(\log(1.3), 0.075^2), \hskip 10pt 
\kappa \sim \mathcal{U}(0.01, 0.05).
\end{align*}
Further details on how the parameters of the above design priors were selected are provided in the Supplementary Material.

% we define the following multivariate normal random variable that represents the joint decision statistic  for an experiment with $T$ analyses:  $$\boldsymbol{\gamma} = \mathcal{N}\left(\dfrac{(\psi^* - \psi_0)}{\hat{\lambda}}\begin{pmatrix}
% \sqrt{n_1}\\
% \sqrt{n_2}\\
% \vdots \\
% \sqrt{n_T}
% \end{pmatrix}, \begin{bmatrix}
% 1 & \sqrt{n_1/n_2} & \dots & \sqrt{n_1/n_T}\\
% \sqrt{n_1/n_2} & 1 & \dots & \sqrt{n_1/n_T}\\
% \vdots & \vdots & \ddots & \vdots \\
% \sqrt{n_1/n_T} & \sqrt{n_2/n_T} & \dots & 1
% \end{bmatrix}\right).$$
% This joint distribution of the decision statistics at all $T$ stages enables computation of cumulative probabilities of stopping up to each analysis as well as power. For example, 
% the probability that the experiment stops for success at analysis $t \le T$ is 
% $$P(\gamma_1 < \Phi^{-1}(u_1), \gamma_2 < \Phi^{-1}(u_2), \dots, \gamma_t \ge \Phi^{-1}(u_t)),$$
% which can be derived from the above multivariate normal sampling distribution by appropriate conditioning.

{\bf Bayesian group sequential design} To demonstrate the utilities of the proposed approach in a hypothetical design exercise for this study we consider a family of Bayesian GSDs that allow for stopping for efficacy or futility at the interim analyses. Countless possibilities arise from the number and spacing of the interim analyses, the total allowable sample size and the efficacy/futility decision thresholds. Relying on the proposed approach, after estimating the Fisher information across the parameter space, one can instantaneously evaluate several operating characteristics for any given design within the above general family of GSDs. The approximation in (\ref{eq:joint.gamma}) is used to estimate cumulative probabilities of stopping up to each analysis. These probabilities can be computed by summing the BART-BvM estimates in (\ref{eqn:power.gsd}) or (\ref{eqn:assurance.gsd}) across the relevant stages of the GSD.

For illustration, we specify the following two designs and present a variety of operating characteristics that can be used to assess these designs.

\begin{enumerate}
    \item[D1] involves three analysis, two interim analyses at $n=350$ and $n=500$, and a final analysis at $n=700$. The decision of stopping for efficacy is considered at each analysis. Specifically, if $P(\beta>0\mid \mathbf{y}_t)>u_t$ where $\mathbf{y}_t$ is the accumulated data up to analysis $t=1, 2, 3$, and $u_t$ are the corresponding decision thresholds which are specified to be conservative at the initial analysis with a slight decrease in subsequent analyses, $\mathbf{u} = (0.99, 0.98, 0.975)$.
    \item[D2] involves four analysis, three interim analyses at $n=300$, $n=400$, and $n=500$, and a final analysis at $n = 600$. The same efficacy decision as in D1 is considered. The decision thresholds remain equally conservative through the interim analyses but decrease at the final analysis, $\mathbf{u} = (0.99, 0.99, 0.99, 0.95)$.
\end{enumerate}

Initial simulations to generate the training set for $\lambda$ are performed using an embedded Gibbs sampler for posterior approximation. The training set is specified as 60 points given by a Latin hypercube design over a subset of the 6-dimensional parameter space that has non-negligible density under the design prior. The probabilistic BART model is then trained over the generated values to predict $\lambda$ for any given set of parameter values and subsequently the probabilities of making decisions conditioning on the given parameters. To evaluate power integrated over a subset or all the model parameters, we use a sample of size 100,000 from the design prior for Monte Carlo integration. 

\begin{figure}[!tb] \centering 
	\begin{subfigure}[b]{0.75\textwidth}
		\centering
		\includegraphics[width=\textwidth]{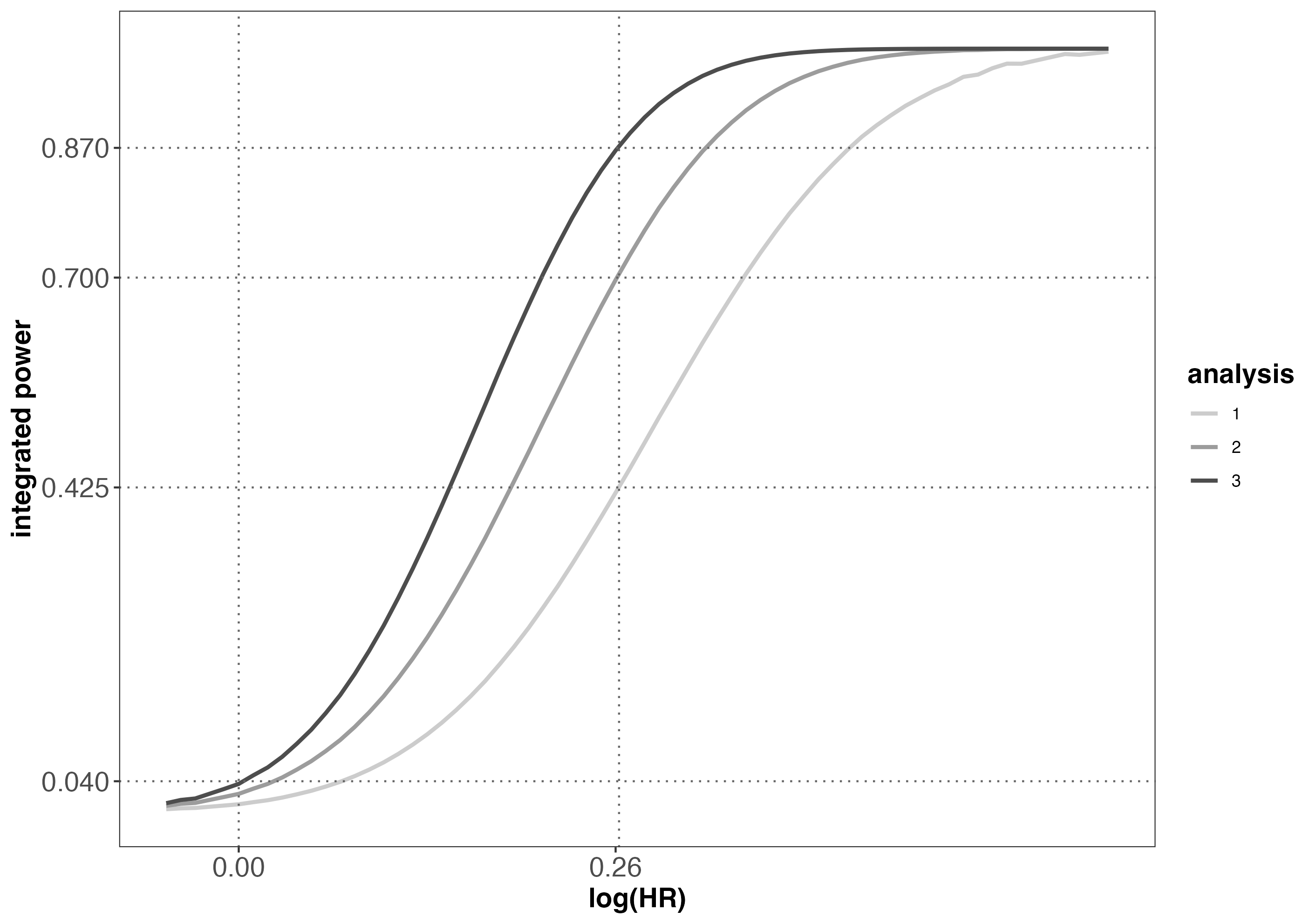}
		\caption{}
		\label{fig:D1}
	\end{subfigure}
    	\begin{subfigure}[b]{0.75\textwidth}
		\centering
		\includegraphics[width=\textwidth]{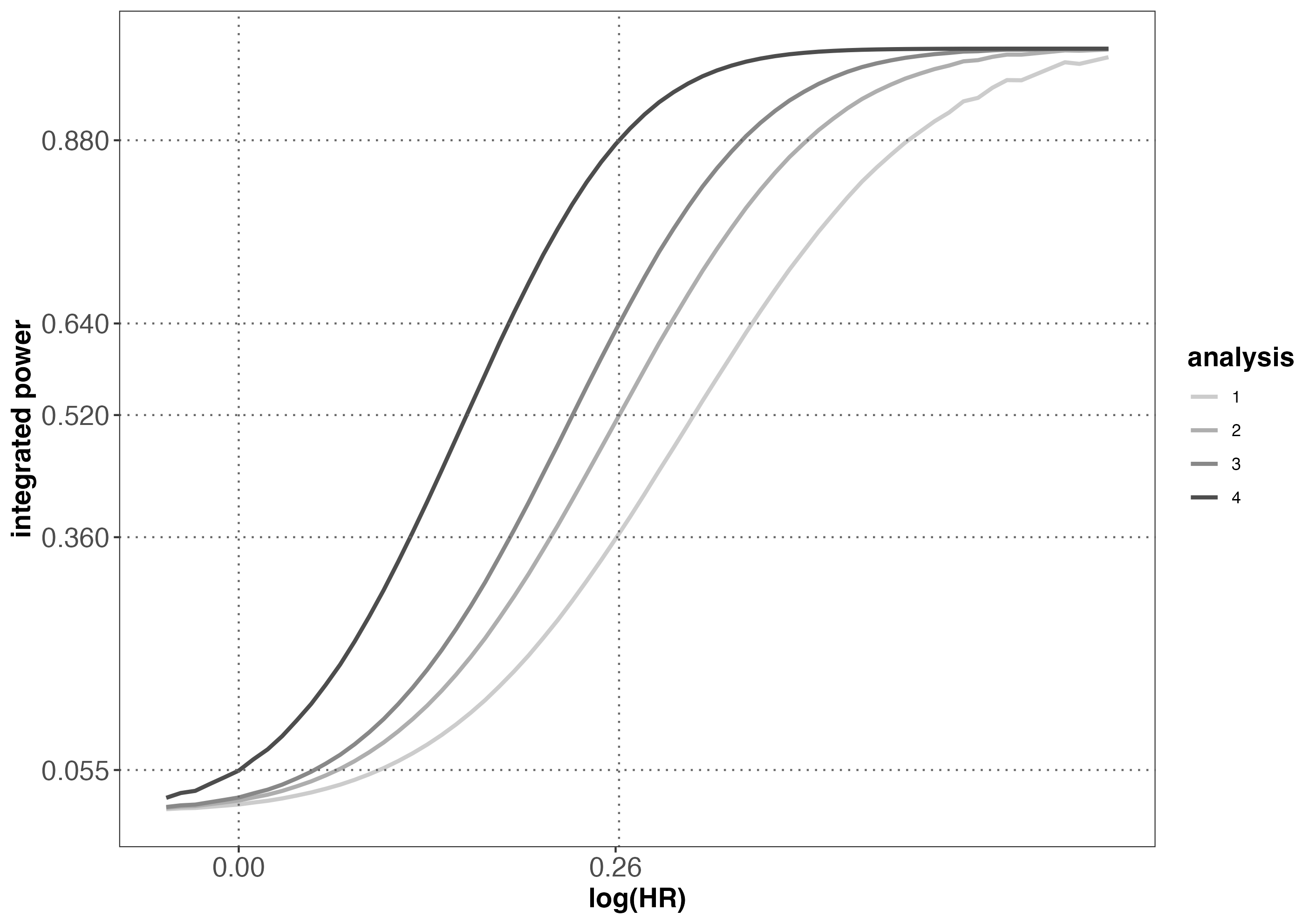}
		\caption{}
		\label{fig:D2}
	\end{subfigure}
\caption{BART-BvM curves for cumulative stopping probabilities integrated over the nuisance parameters versus the log hazard ratio for designs D1 and D2.} 
\label{iPower}
\end{figure}

Figure~\ref{iPower} shows the probabilities of stopping for efficacy at each analysis for a range of values of the treatment coefficient $\beta$ but integrated over the design prior of the nuisance parameters, i.e., the baseline hazards and the censoring parameter. The vertical lines are drawn at the null value, $\beta = 0$ and the most plausible value under the alternative (the mean of the design prior), $\beta=\log(1.3)$. The horizontal lines are drawn to aid in visualizing the cumulative probabilities of stopping for each analysis at these given values for the two designs.

The Bayesian assurance, at each analysis and overall, may be computed by further integrating the integrated (over the nuisance parameters) power function over the design prior of $\beta = \log(\text{HR})$. Likewise, the expected sample sizes for each design over all plausible parameter values and the data distribution may be estimated from the estimated probabilities of stopping at each analysis. 

In addition, to showcase the utility of the proposed approach in a decision theoretic setting, we evaluate the ``integrated expected cost'' (IEC) of the trial under each design based on the following loss function. 
\begin{equation}
L(\boldsymbol{\theta}, d(\mathbf{y}_n)) = c_0 \mathbb{I}(\psi(\boldsymbol{\theta})\leq \psi_0)\mathbb{I}(d(\mathbf{y}_n)=1) + c_1 \mathbb{I}(\psi(\boldsymbol{\theta})> \psi_0)\mathbb{I}(d(\mathbf{y}_n)=0) + c_2N(\boldsymbol{\theta}, d(\mathbf{y})),
\end{equation}
where $\mathbb{I}$ is an identity function, $d(\mathbf{y}_n)=1$ represents the decision of concluding efficacy (according to the rule defined in Section \ref{sec:methods.1} with respect to the posterior probability in (\ref{eqn:pAlt})), and the cost parameters $c_0$, $c_1$, and $c_2$ are the cost of making a type I error, making a type II error, and enrolling a patient, respectively. The notation $N(\boldsymbol{\theta}, d(\mathbf{y}))$ is used to emphasize that the final sample size in a GSD is a random variable that depends on the ``true'' value of the parameter and the data. Integrating the above loss function over the design prior and with respect to the sampling distribution for a given design results in the following function which we refer to as the IEC function.
\begin{equation}
\text{IEC} = c_0 \int_{\Theta_0} \Pi(\boldsymbol{\theta})\pi_d(\boldsymbol{\theta})d\boldsymbol{\theta} + c_1 \int_{\Theta_A} (1-\Pi(\boldsymbol{\theta}))\pi_d(\boldsymbol{\theta})d\boldsymbol{\theta} + c_2\bar{N},
\end{equation}
where $\bar{N} = \sum_{t = 1}^T\Pi_B^tn_t$ is the integrated expected sample size (IESS) that is estimated from the integrated probability of stopping \emph{at} analysis $t = 1, \ldots, T$.

Table~\ref{BAESS} shows the Bayesian power (assurance), IESS and IEC for the two designs with the arbitrarily selected cost parameters, $c_0=1000$ (large penalty for a type I error), $c_1 = 10$, and $c_2=1$.

  \begin{table}[!ht]
\centering
\begin{tabular}{c|cccccc}
 & BA1 & BA2 & BA3 & BA4 & IESS & IEC\\
 \hline
 D1 & 0.44 & 0.67 & 0.81 & NA & 365 & 394\\
 D2 & 0.38 & 0.52 & 0.61 & 0.82 & 341 & 384\\
\end{tabular}
\caption{Operating characteristics of the two designs D1 and D2. BA: Bayesian assurance --cumulative probability of making an efficacy decision at each analysis, IESS: Integrated expected sample size, IEC: Integrated expected cost.}
\label{BAESS}
\end{table}

In this hypothetical design scenario, the two designs result in a similar level of overall Bayesian assurance (about 80\%). D2 can result in potential savings as it has a smaller IESS and IEC, but since D1 is not far worse with respect to these criteria, the choice should ultimately be made considering the feasibility of an additional analysis. It is also straightforward to formally incorporate the cost associated with each analysis into the cost function.

We investigated the quality of the BART predictions for the variance parameter as well as the estimated cumulative power at the three analyses of D1, using leave-one-out cross validation. We note that a larger deviation from Monte Carlo estimates may be observed at certain interim analyses due to deviations of the posterior distribution from the normal approximation. In this example, the maximum observed deviation was under 0.04 which is considered negligible from a practical perspective. These results are presented in the Supplementary Material.

\section{Discussion}
\label{Sec:discussion}

In this manuscript, we have proposed an approach to enable the efficient Bayesian design of experiments in the presence of nuisance parameters. We consider settings where design, analysis and decision making are within the Bayesian framework. Particularly, design priors are specified for all model parameters to incorporate the uncertainty about the nuisance parameters and the hypothesized values of the parameter(s) of interest at the design stage. This is different than the analysis prior that completes the Bayesian analysis model leading to the posterior distribution whose summaries are used for decision making. 

Our proposed approach, BART-BvM, relies on asymptotic properties of the posterior distribution given by the Bernstein-von Mises theorem to obtain the sampling distribution of posterior decision summaries up to the unknown variance parameter. This parameter is learned empirically for a small set of points throughout the parameter space. The empirical estimates are then used to train a probabilistic BART model and predict the variance across the entire, or more practically a plausible subset, of the parameter space. This leads to efficient approximation of  criteria such as Bayesian analogues to frequentist power and expected loss/utility/cost in a decision theoretic framework. 

The proposed approach was implemented and assessed within the context of a stylized example (simple design and analysis setting). We investigated the performance of BART-BvM with respect to how well the BART predictions recovered the variance parameter of the sampling distribution, and how the power estimates resulting from the normal approximation with the predicted variance parameter compared to empirical power estimates. 

The performance of the proposed approach appears highly satisfactory in the examples of this manuscript. However, this satisfactory performance is not broadly generalizable to any design and/or analysis since the BvM approximation may be poor for small sample sizes and in more complex models and the BART predictions may be more sensitive to the size and distribution of the training set depending on the analysis model. We note that even in these cases, the proposed approach can be used as an efficient exploratory method in design comparisons. Monte Carlo simulations can then be used for a few final design options to obtain more precise estimates of the operating characteristics. Alternatively, possible extensions to the proposed approach that can improve the performance include small-sample corrections to the sampling distribution or the use of linear approximations similar to those proposed in \citet{hagar2025scalable, hagar2025design} and \citet{hagar2025sequential} rather than relying on the normal approximation to the posterior distribution.

We emphasize that a strength of the proposed approach is enabling the decision theoretic framework in the design of studies, especially in clinical trials, by diminishing the computational barriers. Here we considered a pragmatic cost function inspired by \citet{PerPer2016_balancing} which tends to unify Bayesian and frequentist approaches toward assessing and optimizing the design. However, other similar loss functions, such as that proposed by \cite{Calderazzo2020} to incorporate estimation accuracy and precision can be accommodated.

While the proposed approach is presented for a set of one-sided hypotheses, we note that it can seamlessly be generalized to interval hypotheses. Therefore, our methods have applicability in a broad range of problems with respect to hypotheses, analysis models and designs.

To conclude, we note that the proposed set of methods in this manuscript can significantly contribute to the uptake of Bayesian study design in practice. By increasing the efficiency of design assessment and comparison with respect to criteria that adequately reflect the uncertainties about all model parameters, the proposed methods enable rigorous and robust practices in the design of experiments.

%%%%%%%%%%%%%%%%%%%%%%%%%%%%%%%%%%%%%%%%%%%%%%
%% Acknowledgements                         %%
%% should be provided in the                %%
%% Acknowledgements section.                %%
%%%%%%%%%%%%%%%%%%%%%%%%%%%%%%%%%%%%%%%%%%%%%%
% \begin{acks}[Acknowledgments]
% The authors would like to thank the anonymous referees, an Associate
% Editor and the Editor for their constructive comments that improved the
% quality of this paper.

% \end{acks}

%%%%%%%%%%%%%%%%%%%%%%%%%%%%%%%%%%%%%%%%%%%%%%
%% Funding information, if any,             %%
%% should be provided in the                %%
%% funding section.                         %%
%%%%%%%%%%%%%%%%%%%%%%%%%%%%%%%%%%%%%%%%%%%%%%
\section*{Code}
The code to reproduce the results in Sections 3 and 4 of the manuscript is available at the following public repository: \href{https://github.com/sgolchi/BartBvM}{https://github.com/sgolchi/BartBvM}.

\section*{Acknowledgements}
SG acknowledges support from the Natural Sciences and Engineering Research Council of Canada (NSERC), Canadian Institute for Statistical Sciences (CANSSI), and Fonds de recherche du Québec - Santé (FRQS) and Fonds de recherche du Québec – Nature et technologies (FRQNT). LH acknowledges support from an NSERC postdoctoral fellowship.

%%%%%%%%%%%%%%%%%%%%%%%%%%%%%%%%%%%%%%%%%%%%%%
%% Supplementary Material, including data   %%
%% sets and code, should be provided in     %%
%% {supplement} environment with title      %%
%% and short description. It cannot be      %%
%% available exclusively as external link.  %%
%% All Supplementary Material must be       %%
%% available to the reader on Project       %%
%% Euclid with the published article.       %%
%%%%%%%%%%%%%%%%%%%%%%%%%%%%%%%%%%%%%%%%%%%%%%
%\begin{supplement}

%\stitle{Additional information for the PLATINUM-CAN example and the code to reproduce the results of the manuscript}

%\sdescription{The supplementary document includes justification of the selected design prior, assessment of BART-BvM for the operating characteristics of D1, as well as an access link to and a brief description of the code to reproduce the results of the manuscript.}
%\end{supplement}

\bibliographystyle{apalike}
\bibliography{refs}
\appendix
\section{Additional information for the PLATINUM-CAN example}
\subsection{Specification of the design prior}
Details on how the parameters of the design priors in Section 4 are selected, and how to tweak these parameters if necessary, are provided as follows. Please note that the following information serves as an example for how  design priors are typically selected and may not precisely reflect all the assumptions and information in the PLATINUM-CAN trial context.

Design priors are assigned to the logarithm of the baseline hazards. The marginal design priors on the logarithms of the baseline hazards ensure that a 95\% ``design'' prior interval on the percentage of control arm patients who do not experience lesion resolution by the end of the 28-day observation period is (17.3\%, 28.5\%). 

The design prior on $\beta$ is centered at a rate ratio of 1.3. This marginal design prior ensures that there is a 95\% design prior interval of (1.123, 1.506) for the rate ratio. In conjunction with the design prior on the logarithms of the baseline hazards, the design prior on $\beta$ gives rise to a 95\% ``design'' prior interval on the percentage of treatment arm patients who do not experience lesion resolution by the end of the 28-day observation period of (9.0\%, 21.4\%). Lastly, the design prior on $\kappa$ ensures that the censoring proportion of patients who have not yet experienced lesion resolution after each clinical visit is relatively low. 

\subsection{Assessment of BART-BvM estimates}

\begin{figure}[!tb] \centering 
	
		\centering
		\includegraphics[width=\textwidth]{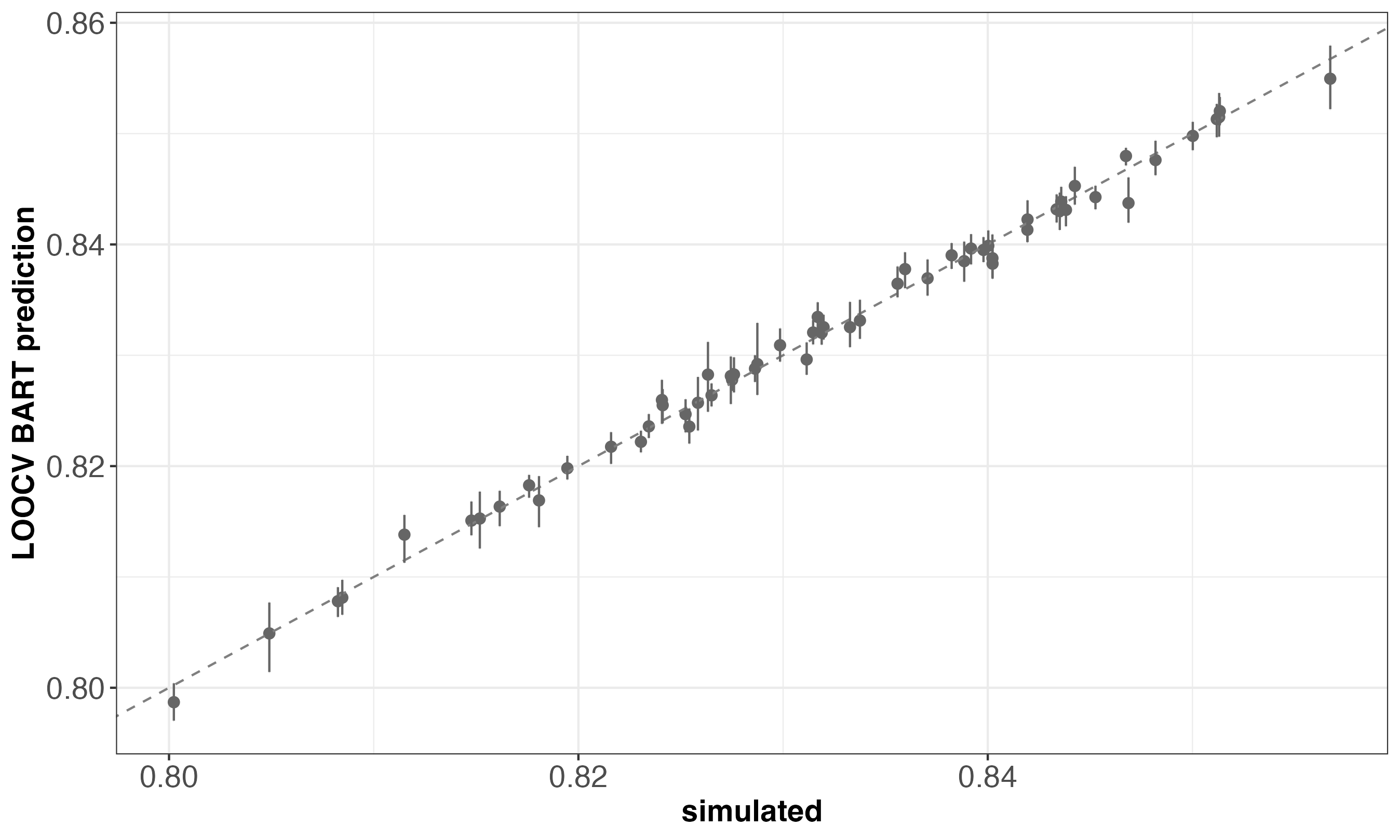}

\caption{LOOCV BART predictions of $\lambda$ versus simulated values on the log scale together with 95\% credible intervals.} 
\label{fig:loocvlambda_PlatCAN}
\end{figure}

\begin{figure}[!tb] \centering 
\begin{subfigure}[b]{0.75\textwidth}
		\centering
		\includegraphics[width=\textwidth]{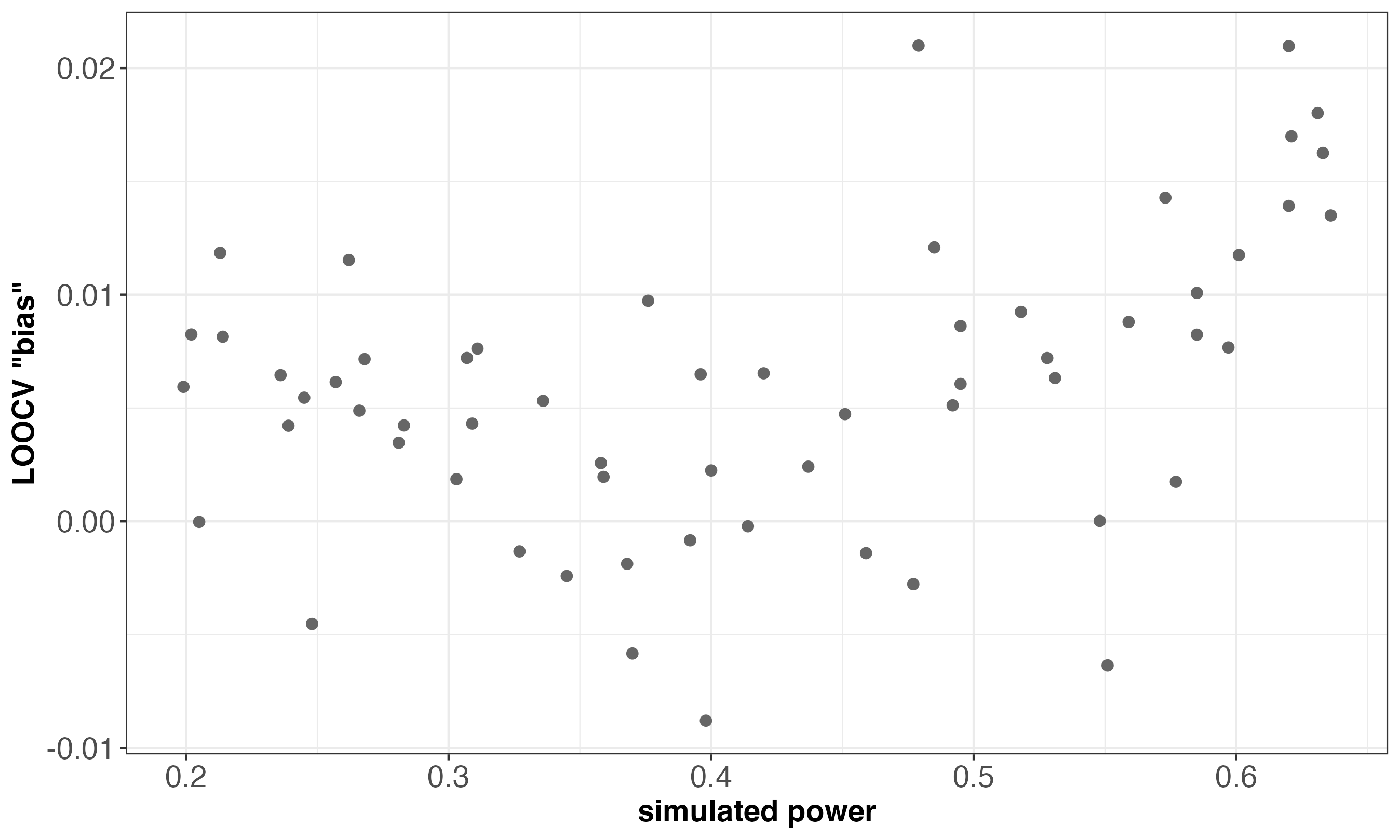}
		\caption{}
		\label{fig:loocvpower1}
	\end{subfigure}
	\begin{subfigure}[b]{0.75\textwidth}
		\centering
		\includegraphics[width=\textwidth]{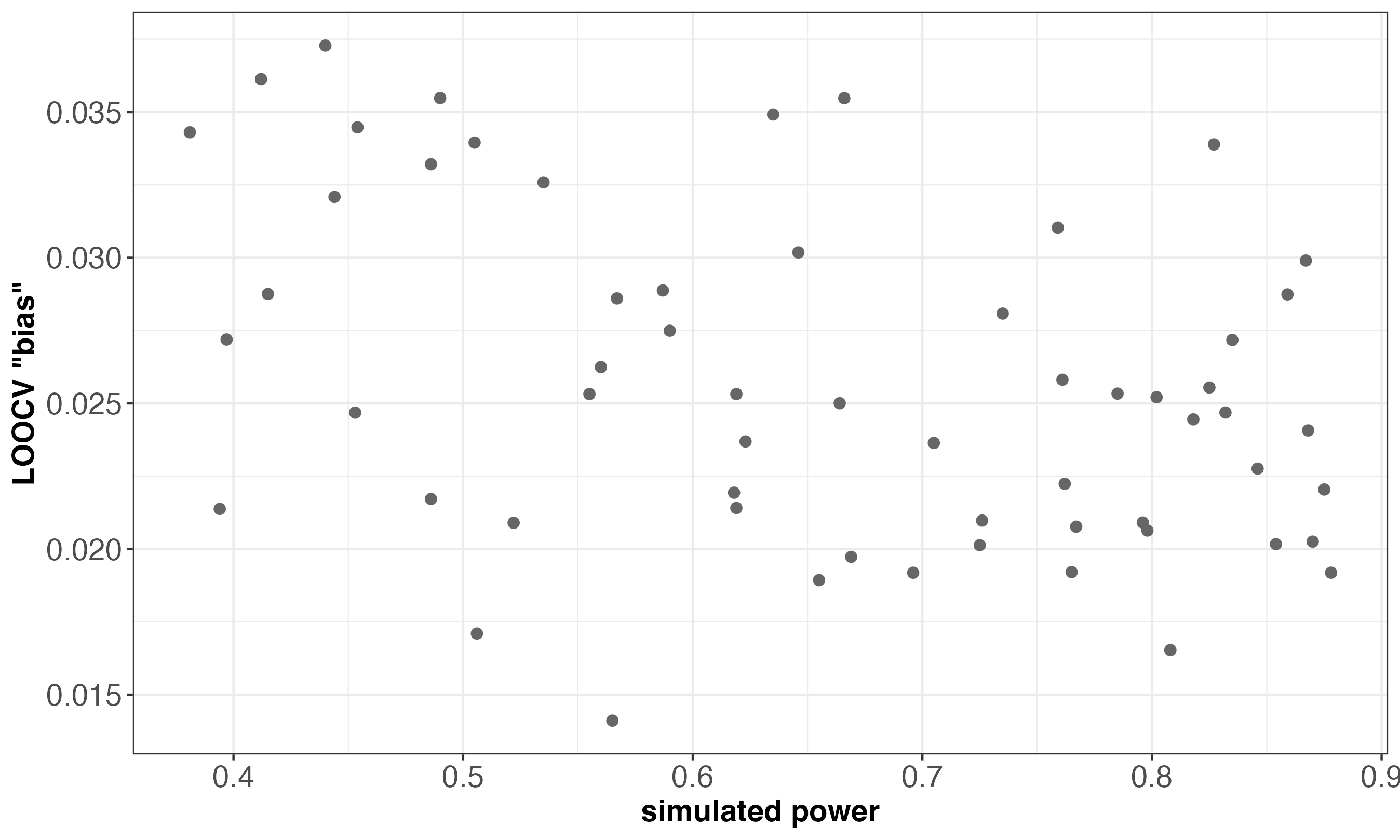}
		\caption{}
		\label{fig:loocvpower2}
	\end{subfigure}
    	\begin{subfigure}[b]{0.75\textwidth}
		\centering
		\includegraphics[width=\textwidth]{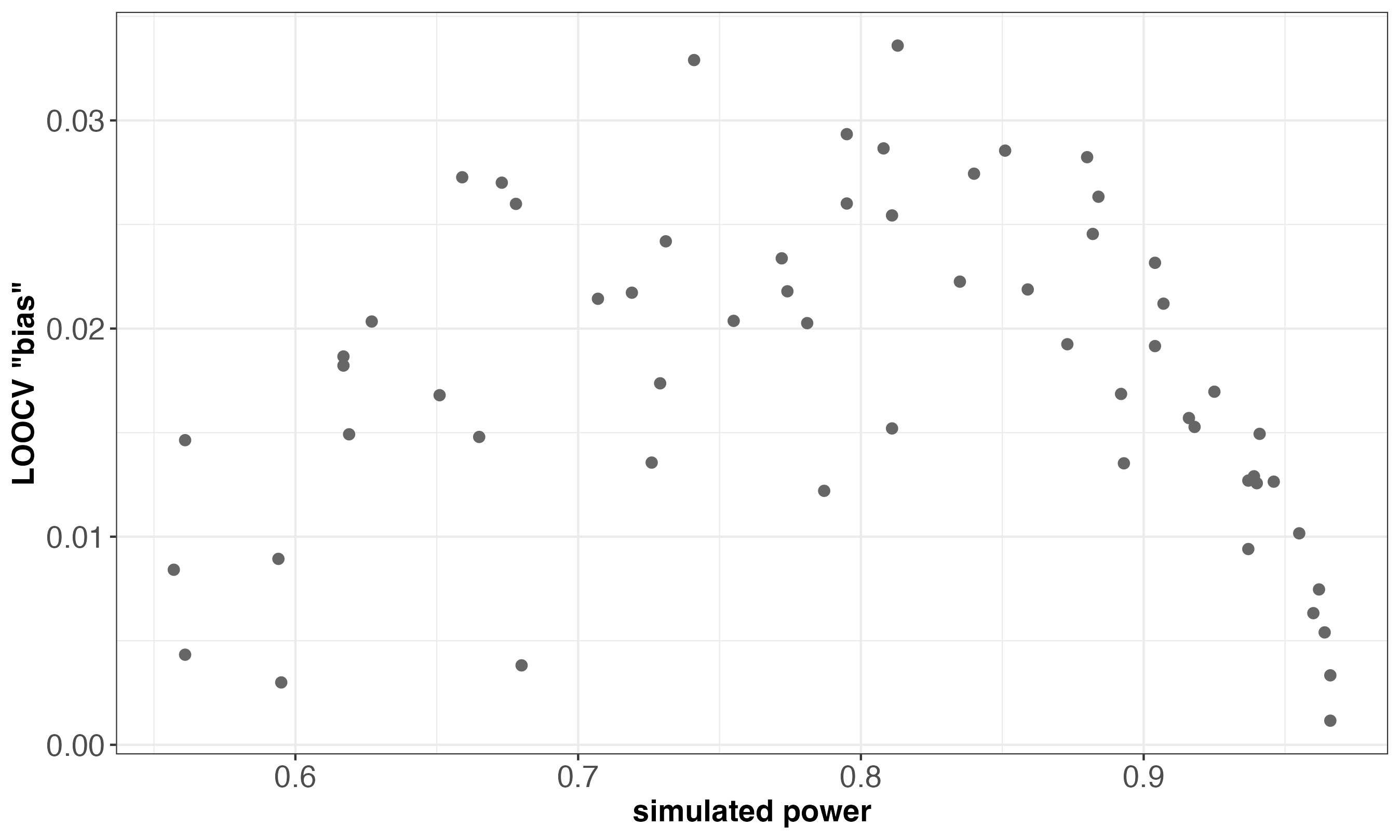}
		\caption{}
		\label{fig:loocvpower3}
	\end{subfigure}
  
\caption{Deviation between BART-BvM estimates and Monte Carlo estimates of cumulative power through the (a) first, $n=350$, (b) second, $n=500$, and (c) last, $n=700$, analyses of D1.} 
  \label{fig:loocvpower_PlatCAN}
\end{figure}

We investigated the performance of the BART-BvM method in the context of one of the hypothetical designs (D1) for the PLATINUM-CAN trial. 
 Figure~\ref{fig:loocvlambda_PlatCAN} shows the leave-one-out cross-validated (LOOCV) predictions generated from the model for the 60 points in the training set. Figure~\ref{fig:loocvpower_PlatCAN} shows the estimates of cumulative power at the three analyses of D1 at $n=350$, $n=500$, and $n=700$, obtained from the BART predictions of $\lambda$ at a single sample size (350) and the BvM approximation compared to those obtained from Monte Carlo simulations. 

\end{document}